\documentclass[12pt,noshowpacs,nofootinbib,notitlepage,amsmath]{revtex4-2}

\usepackage{setspace}
\allowdisplaybreaks
\usepackage{graphicx,color}
\usepackage[charter]{mathdesign}
\usepackage{enumerate}
\usepackage{array}
\usepackage{braket}
\usepackage{slashed}

\begin{document}

\title{Particle creation using the classical stochastic method}

\author{Takayuki Hirayama}
\email[]{hirayama@isc.chubu.ac.jp}
\affiliation{Department of Mathematical and Physical Sciences, 
College of Science and Engineering, Chubu University
\\
1200 Matsumoto, Kasugai, Aichi, 487-8501, Japan}


\begin{abstract}
We compute the particle creation of a harmonic oscillator using the classical stochastic method. This recently constructed method reproduces all the vacuum expectation values in quantum theory. We prepare the vacuum state at the initial time and evolve it over time using Langevin equations of motion. By averaging over the ensemble, we compute the energy of the state at the final time and determine the amount of particles created. We verify that the particle creation agrees with predictions from quantum theory and thus give an evidence that this method can 
really compute the quantum dynamics.
\\[-2ex]


\end{abstract}


\maketitle


\section{Introduction}

The so-called classical statistical method provides a tool for computing real-time 
quantum dynamics. When the occupation number is large, the classical statistical method approximates both quantum and classical theory, as shown in previous studies~\cite{Son:1996zs, Aarts:1997kp, Aarts:2001yn}. When this method is applied to the evolution of the vacuum~\cite{Hirayama:2005ba, Hirayama:2005ac, Gelis:2013oca, Braden:2018tky, Hertzberg:2020tqa, Tranberg:2022noe}\footnote{In~\cite{Hertzberg:2020tqa, Braden:2018tky, Tranberg:2022noe}, $\epsilon$ is introduced in the amplitude of the initial values, and $\epsilon \sim 0.5$ fits the data with the results in quantum theory although $\epsilon=1$ represents the physical value.}, qualitatively good results and the discrepancies
from quantum field theory have been reported. In fact, the discrepancies are expected, as the classical statistical method is known to differ from quantum theory
(see, for example, \cite{Mou:2019gyl, Millington:2020vkg}). 
However, \cite{Hirayama:2019wml} developed a classical statistical method that has been shown to match quantum theory exactly to any order of perturbation.

Thus, it is worth further developing the classical stochastic method described in \cite{Hirayama:2019wml} to enable the calculation of quantum dynamics. 
In this paper, we use this classical stochastic method to calculate a simple yet important quantum dynamic: particle creation in the quantum theory of a harmonic oscillator with a time-dependent frequency. We prove that this method exactly reproduces the predictions from quantum theory.

The classical stochastic method begins with random initial conditions and evolves the system using Langevin equations of motion with complex Gaussian noises. In Section~\ref{sec2}, we explain the classical stochastic method and
demonstrate that both equal-time and unequal-time correlation functions match those in quantum theory, highlighting the differences from the previous classical statistical methods. In Section~\ref{sec3}, we compute particle creation and verify that our results agree with quantum theory. In Section~\ref{sec4}, we summarize our findings.

\section{Classical Stochasitc Method}
\label{sec2}

In this section, we explain the classical stochastic method for computing quantum theory. We apply the method from \cite{Hirayama:2019wml} to quantum mechanics. Specifically, we quantize the Hamiltonian of a harmonic oscillator with mass $m$ and angular frequency $\omega$,
\begin{align}
  H=\frac{p^2}{2m} + \frac{1}{2}m\omega^2 x^2,
  \hspace{3ex}
  (t_i\leq t \leq t_f)
  ,
\end{align}
where $t_i$ denotes the initial time and $t_f$ denotes the final time.
In quantum mechanics, the commutation relation between the operators
 $x$ and $p$ is given by $[x,p]=i\hbar$. Consequently, the equations for $x(t)$ 
 and $p(t)$ in the Heisenberg picture are
\begin{align}
  \dot{x}(t) = \frac{p(t)}{m},
  \hspace{3ex}
  \frac{\dot{p}(t)}{m} +\omega^2 x(t) =0
  .
\end{align}
The solution to these equations is
\begin{align}
  &
  x(t) = \sqrt{\frac{\hbar}{2m\omega}} \Big\{
  a e^{-i\omega t} + a^{\dag} e^{i\omega t}
  \Big\}
  ,
  &
  p(t) = -i \sqrt{\frac{\hbar m\omega}{2}} \Big\{
  a e^{-i\omega t} - a^{\dag} e^{i\omega t}
  \Big\}
  ,
\end{align} 
where $a$ and $a^\dag$ are the annihilation and creation operators
that satisfy the commutation relation $[a, a^\dag]=1$. Then
the energy of the vacuum state $|\Omega\rangle$ 
is $\langle \Omega | H | \Omega \rangle=\hbar \omega/2$,
and T-product $\langle \Omega | \mbox{T} x(t_1) x(t_2) | \Omega \rangle$ is
\begin{align}
  \langle \Omega | \mbox{T} x(t_1) x(t_2) | \Omega \rangle
  &= \frac{\hbar}{2m\omega} \Big[ \theta(t_1-t_2) e^{-i\omega (t_1-t_2)}
  +\theta(t_2-t_1) e^{-i\omega (t_2-t_1)}
  \Big]
  \\
  &=D_F(t_1-t_2)
  .
\end{align}
The Feynman propagator $D_F(t_1-t_2)$,
$(\partial_t^2+\omega^2)D_F(t)=-i(\hbar/m)\delta(t)$,
 is the propagator in quantum theory, while the
retarded propagator is the propagator in classical theory.
We reproduce the Feynman propagator and other arbitrary vacuum 
expectation values (including any arbitrary time-ordered product) using the classical stochastic method.
In the classical stochastic method, the creation and annihilation of particles
are driven by the complex Gaussian noises $J(t)$, and the zero point energy $\hbar \omega/2$ is represented as the energy associated with the white noise background.

  We thus incorporate the noise term into the equation of motion for $x(t)$ and $p(t)$, and the equations which we should solve are
\begin{align}
  \dot{X}(t)&=
  \frac{P(t)}{m}+M~ \Big[\int_{t_i}^{t_f} dt' \theta(t-t') J(t')\Big],
  &
  &\frac{\dot{P}(t)}{m} +\omega^2 X(t)=0.
    \label{e1}
  \\
  x(t)&=X(t)-i\frac{1}{M}J^*(t),
  &
  &p(t) =P(t)-i\dot{J}^*(t),
\end{align}
where $t_i$ and $t_f$ are the initial and final time.
The noises $J(t)$ and $J^*(t)$ are the complex Gaussian noises\footnote{
A complex Gaussian noise $z$ consists of two Gaussian noises $x$ and $y$ as
$z=x+iy$ and $z^*=x-iy$. 
} which satisfy
\begin{align}
  \langle J(t_1) J^*(t_2) \rangle = \frac{\hbar}{2} \frac{\delta(t_1-t_2)}{m}
  .
\end{align}
Here, the braket denotes the average over the ensemble.
The time derivative $\dot{J}(t)$ is defined
\begin{align}
  \dot{J}(t) = \lim_{\Delta t\rightarrow 0+} \frac{J(t)-J(t-\Delta t)}{\Delta t}
  .
\end{align}
We introduced a mass scale $M$ in various places so that $J(t)$ has mass dimension zero. In fact the final results obtained from computing the correlation functions do not depend on $M$.
From \eqref{e1}, we have the equation
\begin{align}
  \ddot{X}(t) +\omega^2 X(t) =MJ(t),
\end{align}
and the classical solution to the equation can be written using
 the retarted propagator,
\begin{align}
  X(t) &= \sqrt{\frac{1}{2m\omega}} \Big\{
  b e^{-i\omega t} + b^* e^{i\omega t}
  \Big\}
 +M\Big[
  \int_{t_i}^{t_f} dt' \theta(t-t') \frac{\sin\big(\omega(t-t')\big)}{\omega} J(t')
  \Big]
  .
\end{align}
Then we have
\begin{align}
  x(t) &= \sqrt{\frac{1}{2m\omega}} \Big\{
  b e^{-i\omega t} + b^* e^{i\omega t}
  \Big\}
 +M\Big[
  \int_{t_i}^{t_f} dt' \theta(t-t') \frac{\sin\big(\omega(t-t')\big)}{\omega} J(t')
  \Big]
  -i\frac{1}{M}J^*(t),
  \\
  p(t) &= -i\sqrt{\frac{m\omega}{2}} \Big\{
  b e^{-i\omega t} - b^* e^{i\omega t}
  \Big\}
  +mM \Big[
  \int_{t_i}^{t_f} dt' \theta(t-t') \big\{ \cos\big(\omega(t-t')\big)-1\big\}
  J(t')
  \Big]
  -i\frac{m}{M}\dot{J}^*(t),
\end{align}
where $b$ and $b^*$ are determined by the intial condition at $t=t_i$, specifically 
$x(t_i)$ and $\dot{x}(t_i)$.
In the classical stochastic method, the Gaussian random initial conditions are 
provided, making $b$ and $b^*$ are complex Gaussian noises with
$\langle bb^* \rangle = \hbar/2$.

The correlation functions are computed as the averages over the ensemble 
of Gaussian noises $b$ and $J(t)$. For example,
\begin{align}
  \langle x(t_1)x(t_2) \rangle =& \,
  \frac{1}{2m\omega} \langle bb^*\rangle \big\{
  e^{-i\omega (t_1-t_2)} + e^{i\omega (t_1-t_2)}\big\}
  \nonumber
  \\
  & -i\int_{t_i}^{t_f} dt' 
  \theta(t_1-t') \frac{\sin\big(\omega(t_1-t')\big)}{\omega}\langle J(t')J^*(t_2) \rangle
  \nonumber
  \\
  & -i\int_{t_i}^{t_f} dt'
  \theta(t_2-t') \frac{\sin\big(\omega(t_2-t')\big)}{\omega}
  \langle J(t')J^*(t_1) \rangle
  \\
  =&\,
  \frac{1}{2m\omega} \frac{\hbar}{2} \big\{
  e^{-i\omega (t_1-t_2)} + e^{i\omega (t_1-t_2)}\big\}
  -i\frac{\hbar}{2} \theta(t_1-t_2) \frac{\sin\big(\omega(t_1-t_2)\big)}{m\omega}
  \nonumber
  \\
  &
  -i\frac{\hbar}{2} \theta(t_2-t_1) \frac{\sin\big(\omega(t_2-t_1)\big)}{m\omega}
  = D_F(x_1-x_2)
  ,\label{14}
  \\
  \langle H \rangle =&\, \langle \frac{p^2}{2m} + \frac{1}{2}m\omega^2 x^2 \rangle
  = \frac{\hbar\omega}{4} +\frac{\hbar\omega}{4}= \frac{\hbar\omega}{2}
  .
\end{align}
Here it is evident that the noises $J(t)$ should be complex noises
rather than real noises in order to reproduce the Feynman propagator. In contrast,
the previous classical statistical methods did not include the complex noises;
only the real noises, if any, were incorporated. We also note that the noises
$J(t)$ do not wash out the initial conditions, which reamain as the first term
in \eqref{14}.

Other correlation functions are also shown to agree with those in quantum theory. For example $\langle x(t_1)x(t_2)x(t_3)x(t_4)\rangle$ matches with
$\langle\Omega|\mbox{T}x(t_1)x(t_2)x(t_3)x(t_4)|\Omega \rangle$ in quantum theory. In this computation the property in Gaussian distribution is importatnt,
\begin{align}
  \langle J(t_1) J(t_2) J^*(t_3)J^*(t_4) \rangle
  =
  \langle J(t_1) J^*(t_3) \rangle
  \langle J(t_2) J^*(t_4) \rangle
  +
  \langle J(t_1) J^*(t_4) \rangle
  \langle J(t_2) J^*(t_3) \rangle
  .
\end{align}

\section{Particle creation}
\label{sec3}

In this section, we demonstrate particle creation using the classical stochastic 
method. We investigate the quantum mechanics of a harmonic oscillator with
mass $m$ and a time-dependent angular frequency $\omega(t)$ which
changes from $\omega$ to $\omega'$ at $t=0$.  The Hamiltonian is
\begin{align}
  H 
  =\frac{p^2}{2m} +\frac{1}{2}m\omega^2(t) x^2
  ,\hspace{3ex}
  \omega(t)
  &= \left\{\begin{array}{lcl}
  \omega,
  &\hspace{3ex}&
  \mbox{for }t_i\leq t\leq 0
  \\[2ex]
  \omega',
  &&
  \mbox{for }t_f\geq t> 0
  \end{array}\right.
  .
\end{align}
In the classical stochastic method, the solution to the equations of 
motion~\eqref{e1} is given by
\begin{align}
  x(t)= \sqrt{\frac{1}{2m\omega}} \Big\{
  b e^{-i\omega t} + b^* e^{i\omega t}
  \Big\}
 +M\Big[
  \int_{t_i}^{0} dt' \theta(t-t') \frac{\sin\big(\omega(t-t')\big)}{\omega} J(t')
  \Big]
  -\frac{i}{M}J^*(t),
\end{align}
for $t\leq 0$ and for $t>0$,
\begin{align}
  x(t)= \sqrt{\frac{1}{2m\omega'}} \Big\{
  b' e^{-i\omega' t} + b'^* e^{i\omega' t}
  \Big\}
 +M\Big[
  \int_{0}^{t_f} dt' \theta(t-t') \frac{\sin\big(\omega'(t-t')\big)}{\omega'} J(t')
  \Big]
  -\frac{i}{M}J^*(t).
\end{align}
$b$ and $b^*$ are determined from the initial condition and are taken as Gaussian white noise with the average $\langle bb^*\rangle=\hbar/2$.
$b'$ and $b'^*$ are determined from the continuity conditions at 
$t=0$, ensuring the continuity of $X(t)$ and $P(t)$.
\begin{align}
  X(0-)&=X(0+)
  &
  &
  \rightarrow
  &
  \sqrt{\frac{1}{2m\omega}} (b+b^*) +\Delta &= 
  \sqrt{\frac{1}{2m\omega'}} (b'+b'^*)  
  ,
  \\
  P(0-)&= P(0+)
  &
  &
  \rightarrow 
  &
  -i\sqrt{\frac{m\omega}{2}} (b-b^*) +\dot{\Delta}&= 
  -i\sqrt{\frac{m\omega'}{2}} (b'-b'^*)
  ,
\end{align}
where
\begin{align}
  \Delta &=
  -M\int_{t_i}^{0} dt' \frac{\sin (\omega t')}{\omega} J(t')
  ,
  &
  \dot{\Delta} &=
  mM\int_{t_i}^{0} dt' \big\{\cos (\omega t' ) -1\big\}J(t')
  .
\end{align}
Then we have,
\begin{align}
  b' &= \alpha b - \beta b^* +\delta
  ,
  &
  \alpha &= \frac{1}{2} \Big\{
  \sqrt{\frac{\omega}{\omega'}}+ \sqrt{\frac{\omega'}{\omega}}\Big\}
  ,
  \\
  b'^* &= -\beta b + \alpha b^* +\delta^*
  ,
  &
  \beta &= \frac{1}{2} \Big\{
  \sqrt{\frac{\omega}{\omega'}} - \sqrt{\frac{\omega'}{\omega}}\Big\}
  ,
\end{align}
and
\begin{align}
  \delta &=
  \sqrt{\frac{m\omega'}{2}}( \Delta +\frac{\dot{\Delta}}{\omega'} )
  ,
  &
  \delta^* &= 
  \sqrt{\frac{m\omega'}{2}}( \Delta -\frac{\dot{\Delta}}{\omega'} )
  .
\end{align}
$\alpha$ and $\beta$ are called the Bogoliubov coefficients, which satisfy
$|\alpha|^2-|\beta|^2=1$. 

We compute the average of Hamiltonian over the noises at $t>0$. In the Hamiltonian with $t>0$, only $J^*(t)$ with $t>0$ appears, and $J^*(t)$ 
with $t\leq 0$ does not appear. Consequently, in computing the average 
over the noises, since $J(t)$ with $t>0$ is not included in $\delta$, 
$\delta$ does not contribute to the average. 
Since,
\begin{align}
  \langle b'b'^\dag \rangle &= \frac{\hbar}{2} (|\alpha|^2 +|\beta|^2)
  =\frac{\hbar}{2} +\hbar|\beta|^2
  ,
\end{align}
the Energy for $t>0$ is computed as
\begin{align}
  \langle H \rangle =
  \langle \frac{p^2}{2m} +\frac{1}{2}m\omega'^2x^2 \rangle 
  = \frac{\hbar \omega'}{2} +\hbar \omega' |\beta|^2
  .
\end{align}
Thus the number of the particle created $N$ is
\begin{align}
  N= |\beta|^2
  =\frac{(\omega-\omega')^2}{4\omega\omega'}
  .
\end{align}
This is exactly the same as in quantum theory.


\section{Summary and Discussion}
\label{sec4}

Previous classical statistical methods were not accurate and were merely approximations of quantum theory. Results using these methods either provided qualitatively correct calculations of quantum dynamics or led to incorrect results. In contrast, the classical stochastic method matches the predictions of quantum theory precisely. It is crucial to further develop this method for computing quantum dynamics and to provide supporting evidence. In this paper, we discussed particle creation in quantum theory using the classical stochastic method and demonstrated that it can accurately compute real-time dynamics.
Therefore, additional calculations of various real-time quantum dynamics using this method are warranted. For example, the particle creation discussed in this paper incorporates the essential elements of Hawking radiation. Consequently, applying this method to the computation of Hawking radiation may help to elucidate the puzzle of the information loss problem and quantum entanglement.
Furthermore, this method also allows for the computation of interacting theories, which is currently a work in progress.

\begin{acknowledgments}
The author would like to thank Koichi Yoshioka for discussion.
\end{acknowledgments}

\appendix

\section{continuous Gaussian noise}

We prepare the real Gaussian noises $J_R(t)$ as the continuous limit of the real gaussian noises $j(t_i)$ in a discrete space. We control the mass dimension using
a mass scale $M$ and ensure that the mass dimensions of the noises are zero.

For $t\in [t_i,t_{i+1}=t_i+\Delta t]$,
\begin{align}
 J_R(t) = \lim_{\Delta t\rightarrow 0} \frac{j(t_i)}{\sqrt{M \Delta t}}
\end{align}
$j(t_i)$ are the gaussian noise $\mbox{N}(0,\sigma^2)$ with the variance $\sigma^2$,
and then $\frac{j(t_i)}{\sqrt{M \Delta t}}$ obeys 
$\mbox{N}(0,\sigma^2/(M\Delta t))$.
Taking this limit, we have
\begin{align}
  \langle J_R(t_1) J_R(t_2) \rangle = \sigma^2 \frac{\delta(t_1-t_2)}{M}
  ,
\end{align}
since
\begin{align}
  \langle \int\!\! dt_2 J_R(t_1) J_R(t_2) \rangle = 
  \langle \lim_{\Delta t\rightarrow 0}  \sum_j \Delta  t\frac{ j(t_i) j(t_j)}{M \Delta t}
  \rangle
  = \frac{\sigma^2}{M}
  .
\end{align}
We compute
\begin{align}
  W(t) = M\!\int^t\!\! J(t')dt' = M \lim_{\Delta t\rightarrow 0}
  \sum_i \Delta t \frac{j(t_i)}{\sqrt{M\Delta t}}
  .
\end{align}
Then $W(t)-W(s)$ obeys $\mbox{N}(0,M(t-s)\sigma^2)$ and
$dW(t)=W(t_i)-W(t_i-1)$.
We define
\begin{align}
  \dot{J}(t) = \lim_{\Delta t\rightarrow 0+} \frac{J(t)-J(t-\Delta t)}{\Delta t}
  ,
\end{align}
and, for example, we have
\begin{align}
  \langle \! \int\!\! dt' F(t')J(t') \dot{J}(t) \rangle
  &=
  \langle \lim_{\Delta t\rightarrow 0+} 
  \! \int\!\! dt' F(t')J(t') 
  \frac{J(t)-J(t-\Delta t)}{\Delta t} \rangle
  \\
  &
  =
  \lim_{\Delta t\rightarrow 0+} \sigma^2
  \frac{F(t)-F(t-\Delta)}{\Delta t}
  =\sigma^2 \partial_t F(t)
  .
\end{align}

\end{document}